[To Appear in Optics Express]

# Debiased Opto-electronic Joint Transform Correlator for Enhanced Real-Time Pattern Recognition

JULIAN GAMBOA,[1,*] XI SHEN,[1] TABASSOM HAMIDFAR,[1] SHAMIMA, A. MITU,[1] SELIM M. SHAHRIAR[1,2,3]

[1]*Department of ECE, Northwestern University, Evanston, IL 60208, USA*
[2]*Department of Physics and Astronomy, Northwestern University, Evanston, IL 60208, USA*
[3]*shahriar@northwestern.edu*
[*]*JulianGamboa2023@u.northwestern.edu*

**Abstract:** Opto-electronic joint transform correlators (OJTCs) use a focal plane array (FPA) to detect the joint power spectrum (JPS) of two input images, projecting it onto a spatial light modulator (SLM) to be optically Fourier transformed. The JPS is composed of two self-intensities and two conjugate-products, where only the latter produce the cross-correlation. However, the self-intensity terms are typically much stronger than the conjugate-products, producing a bias that consumes most of the available bit-depth on the FPA and SLM. Here we propose and demonstrate, through simulation and experiment, a debiased OJTC (DOJTC) that electronically pre-processes the JPS to remove the self-intensity terms before sending it to the SLM, thereby enhancing the quality of the cross-correlation result. We show that under some conditions the DOJTC yields a nearly two orders of magnitude improvement in the signal-to-noise ratio compared to an OJTC.

## 1. Introduction

Recently, there has been a drastic increase in the amount of data that is generated each day, largely thanks to advances in social media, cameras, and computing. A significant portion of this data is composed of images, figures, or other 2D structures which are often ill-suited for traditional signal processing techniques. For this reason, specialized methodologies have been developed to extract or identify arbitrarily defined patterns in these structures. Techniques based on feature extraction via machine learning (ML), for example, have recently achieved processing speeds on the order of 120 operations per second [1,2]. These types of techniques, however, are attempting to bypass the fundamental challenge posed by the fact that modern digital electronic systems are not optimized for 2D data processing. This challenge can be overcome by using inherently 2D systems to build alternative data processing techniques. Fourier optics is one such domain [3–5], wherein lenses are used to perform the 2D Fourier Transform (FT) of the transverse spatial profile of a laser beam. This profile can be configured using a Spatial Light Modulator (SLM) to contain arbitrary 2D data, limited only by the spatial resolution, bit-depth, and refresh rate of the device. Holographic optical storage devices with very low latency can be used for storing and retrieving images from a reference database in order to avoid some of these constraints, albeit with limitations for some applications [6–9].

An optical correlator or convolver can be constructed by first performing the optical FT of two images, multiplying them together, and then FT'ing the result. Various techniques can be used to generate such products. One approach is the use of the Vander-Lugt correlator, which requires constructing a filter by interfering the FT of the desired reference image with a plane-wave inside a photosensitive medium, thus forming a hologram of the FT [10]. This holographic filter is placed at the output Fourier plane of a lens, with the query image being projected at its corresponding input plane. The diffraction from the filter thus corresponds to the product between the reference and query FTs, which can be optically FT'd using a second lens to produce both the cross-correlation and convolution at the output. This technique is impractical due to the two-step process stemming from the need to replace the holographic filter for each operation. An alternative technique, the Joint Transform Correlator (JTC), instead simultaneously projects both the reference, $R$, and query, $Q$ images at the input plane of a single lens, generating both FTs, $\tilde{R}$ and $\tilde{Q}$, respectively, at the opposite focal plane [3,11]. Because the signals are optical, they interfere with each other to yield a new signal $\tilde{R} + \tilde{Q}$. A nonlinear medium is placed at this Fourier plane, generating the Joint Power Spectrum (JPS):

$$\left|\tilde{R} + \tilde{Q}\right|^2 = \left|\tilde{R}\right|^2 + \left|\tilde{Q}\right|^2 + \tilde{R}^*\tilde{Q} + \tilde{R}\tilde{Q}^* \tag{1}$$

This is subsequently FT'd by another lens, where the latter two terms produce conjugate cross-correlation outputs. Here, the two intensity terms in the JPS, $|\tilde{R}|^2$ and $|\tilde{Q}|^2$, are unavoidably FT'd alongside the two conjugate-product terms, $\tilde{R}^*\tilde{Q}$ and $\tilde{R}\tilde{Q}^*$, producing a bias that reduces the overall quality of the final output [12,13]. Additionally, depending on the exact projection geometry, it is possible for the FT of the independent intensity terms to overlap the FT of the conjugate-product terms, thus making it impossible to extract any meaningful data. This can be avoided by shifting the images relative to each other so as to avoid spatial overlap, effectively forming an off-axis interference pattern [14].

An alternative architecture known as the Hybrid Opto-electronic Correlator (HOC) [15,16] can overcome the issue of the independent intensity terms by separately interfering the FT of each image with their own reference plane wave, labelled $C_r$ and $C_q$, whose relative phases can be controlled. The resulting interference produces the signals $|\tilde{R} + C_r|^2$ and $|\tilde{Q} + C_q|^2$. The intensity of each signal is also simultaneously measured: $|\tilde{R}|^2$, $|\tilde{Q}|^2$, $|C_r|^2$, and $|C_q|^2$. These six signals are then arithmetically combined to remove the intensity terms and produce four product terms, yielding a signal $S$ which can be projected on an SLM to be optically FT'd:

$$\begin{aligned} S &= \left(\left|\tilde{R} + C_r\right|^2 - \left|\tilde{R}\right|^2 - |C_r|^2\right)\left(\left|\tilde{Q} + C_q\right|^2 - \left|\tilde{Q}\right|^2 - \left|C_q\right|^2\right) \\ &= \left(\tilde{R}C_r^* + \tilde{R}^*C_r\right)\left(\tilde{Q}C_q^* + \tilde{Q}^*C_q\right) \\ &= \tilde{R}\tilde{Q}C_r^*C_q^* + \tilde{R}\tilde{Q}^*C_r^*C_q + \tilde{R}^*\tilde{Q}C_rC_q^* + \tilde{R}^*\tilde{Q}^*C_rC_q \end{aligned} \tag{2}$$

If the plane waves are spatially constant, after the final optical FT, these four terms will exactly yield two copies of the convolution and two copies of the cross-correlation of the two input images [14,15]. The arithmetic processing step is performed in the electronic domain, and can be done on a microcontroller, CPU, FPGA, or custom integrated circuits [16]. To date, all experimental demonstrations of the HOC architecture have been performed using a computer for simplicity. However, the required processing is relatively simple and does not require a full-fledged computer.

In the case of an opto-electronic JTC (OJTC), a focal plane array (FPA) replaces the nonlinear medium, detecting the JPS and transmitting it to a computer which subsequently projects it on an SLM to be optically FT'd. Here, the electronic stage acts as a relay between the FPA and the SLM, and as such does not perform any additional processing. As we will discuss in the next section, the limited bit-depth of commercially available SLMs severely hinders this functionality. Here, we propose a debiased OJTC (DOJTC) that combines elements from the OJTC and the HOC to overcome this issue and yield enhanced high-speed pattern recognition. Furthermore, the speed and resolution of an SLM is directly tied to its bit-depth. Currently, 8-bit bit-depth SLMs are available with monochrome framerates on the order of 720 frames per second (fps) at a resolution of 1920 x 1080 pixels [17,18]. In contrast, commercial high-speed FPAs can achieve beyond 5,000 fps at the same resolution and bit-depth [19]. As such, the use of SLMs will introduce a significant bottleneck in the operating speed of any opto-electronic correlator. However, some SLM architectures, such as those relying on digital micro-mirror devices (DMDs), can increase their speed by reducing their bit-depth. The debiasing technique proposed here could be used to mitigate the resulting loss of resolution, enabling opto-electronic correlators to operate in the microsecond regime.

We have previously investigated the HOC architecture and how it can perform shift, scale, and rotation invariant correlation [15,20], as well as optimizing its design to remove the strong dependence on the optical phase of its reference plane waves [14]. However, this device is more complex than the traditional OJTC, making it impractical for certain applications. The DOJTC takes inspiration from the mathematical formulation of the HOC to enhance the OJTC, thereby enhancing its peak output correlation power.

The rest of the paper is organized as follows. Section 2 describes the DOJTC architecture and the corresponding mathematical derivation, as well as simulated outputs. The experimental implementation, results, and a discussion on the implications of the data follow in Section 3, and Section 4 concludes with a summary and outlook.

## 2. Debiased Opto-electronic JTC Architecture

As mentioned above, the OJTC uses an FPA to replace the nonlinear medium. This is illustrated schematically in Fig. 1. In this configuration, the device is separated into two independent FT segments, one functioning as the input stage and the other as the output stage. In the example shown below, each stage is identically composed of one projection SLM, one FT lens, and one FPA. The input SLM simultaneously projects the reference and query images, labelled $R$ and $Q$, respectively, spatially offset from each other by a fixed amount so as to avoid overlapping. The JPS is detected by the input FPA, converting the result into an electronic signal that is transmitted to the output SLM to be projected back into the optical domain. Finally, the output FPA detects the FT of the JPS, yielding the cross-correlation of the two input images.

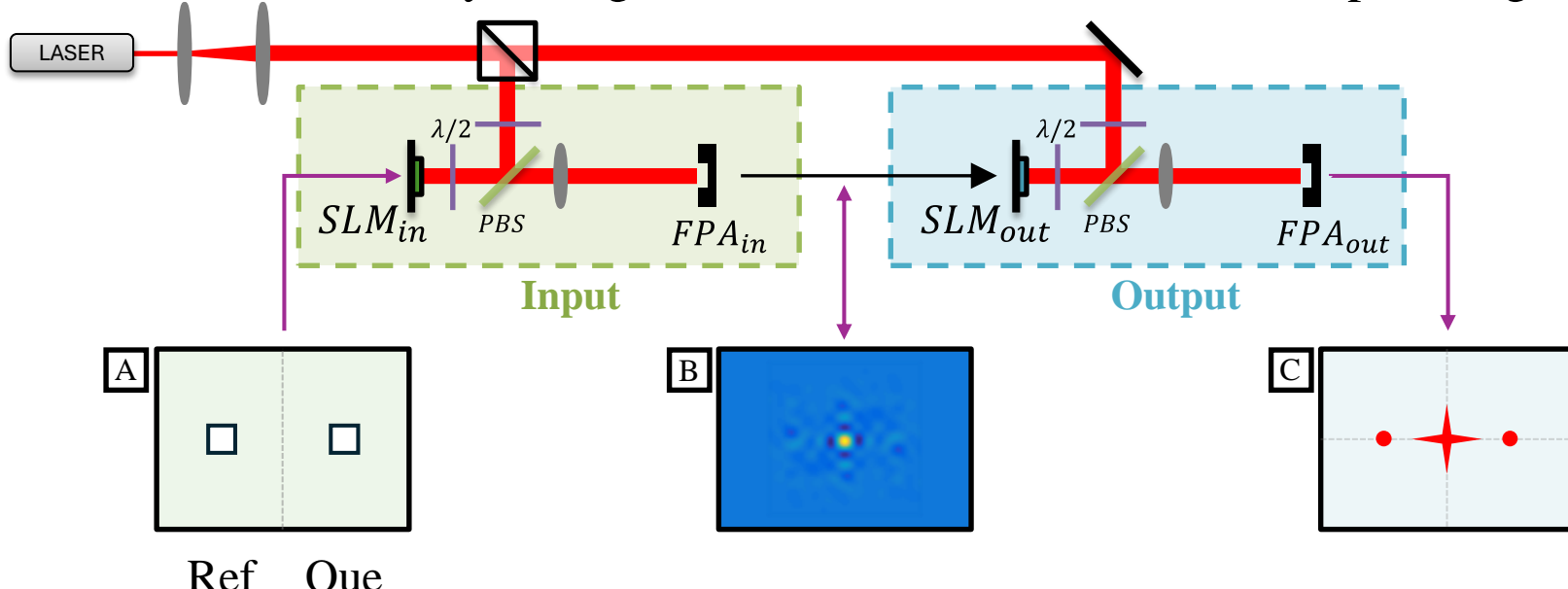

Fig. 1. An example of a classical OJTC architecture. (A): Signal projected onto the input SLM consisting of the reference and query images. (B): JPS Detected by the input FPA and projected onto the output SLM. (C): Final output consisting of two cross-correlation terms at the sides and a large DC term in the center.

The JPS has four components: two independent intensity terms and two conjugate product terms, as shown in eqn. (1). Despite the fact that only the latter are required for the cross-correlation, the system cannot separate out these signals, and so the former must also inevitably be detected on the input FPA and projected on the output SLM. Additionally, the magnitude of the intensity terms will always be equal to or greater than the conjugate product terms. Unfortunately, both FPAs and SLMs possess a limited bit-depth, and so can only detect and project signals with a finite resolution. Thus, if the JPS is detected and projected without further processing, the resolution-space will be significantly biased towards the intensity terms, limiting the resolution of the conjugate product terms, as illustrated in Fig. 2.

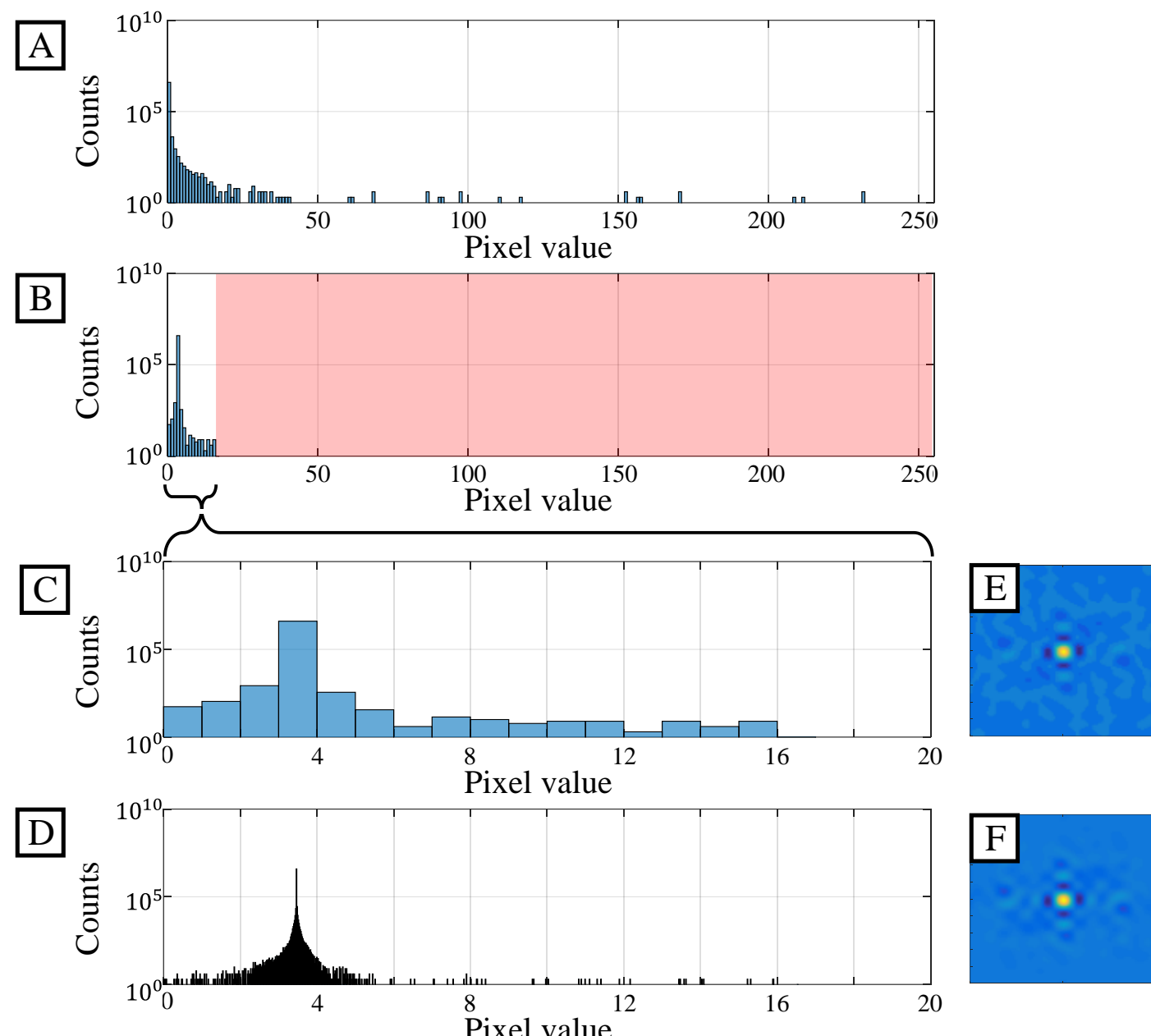

Fig. 2. Simulated Histograms of the integer pixel values that would be sent to the output SLM in an OJTC for a particular cross-correlation. The values are rescaled such that the JPS is between 0-255 inclusive, representing an SLM with a bit-depth of 8-bits. (A): The full JPS, as is used in an OJTC. (B): The conjugate product terms. (C): Closeup of (B). (D): The non-integer data used to construct (C), showing the loss of resolution. (E): 2D data that is sent to the SLM corresponding to the integer pixel values shown in (C). Ideal 2D data from the conjugate product terms, corresponding to the values shown in (D).

Fig. 2 shows simulated histograms of the pixel values that are sent to an SLM for an arbitrary cross-correlation. Fig. 2(A) shows the histogram corresponding to the raw JPS, as used in an OJTC, after rescaling to a maximum integer value of 255 (8-bits). Fig. 2(B) shows the values that correspond only to the conjugate product terms from the same data. These terms include negative values, and so the data has been offset such that the minimum value is zero. The red area highlights values that are taken up exclusively by the intensity terms, and thus represents wasted bit-depth. In this particular case, the useful data has a maximum pixel value of 16, representing only ~6.6% of the available SLM resolution. Fig. 2(C) is a closeup of Fig. 2(B). Fig. 2(D) shows the ideal, non-integer data used to construct Fig. 2(C), wherein it is clear that the integer binning has resulted in a significant loss of information. This is further illustrated in Fig. 2(E) and Fig. 2(F), which show the 2D data corresponding to the histograms in Fig. 2(C) and Fig. 2(D), respectively. Here, the low-resolution integer binning in the former creates a clear banding effect that is not present in the latter. In order to avoid these issues, we propose a debiasing technique to reduce the impact of the two intensity terms, thus optimizing the available resolution for the conjugate product terms, as described next.

### *2.1 DOJTC Design*

Debiasing the JPS requires the removal of the two independent intensity terms in eqn. (1). This can be done as in the HOC in eqn. (2). First, three signals are separately measured: the JPS, the intensity of the FT of the reference image, and the intensity of the FT of the query image. Then, the two intensity terms are subtracted from the JPS, leaving only the two conjugate product terms:

$$S = \tilde{R}^{*}\tilde{Q} + \tilde{R}\tilde{Q}^{*} = \left|\tilde{R} + \tilde{Q}\right|^{2} - \left|\tilde{R}\right|^{2} - \left|\tilde{Q}\right|^{2} \tag{3}$$

This operation can be performed in the electronic domain without incurring significant time-cost. However, this new signal, $S$, may contain negative values, and so it must be rescaled to make it definite-positive. This can also be performed at high speed if the scaling is pre-defined, as discussed below.

As with all opto-electronic correlators, the DOJTC can only operate on targets within a fixed range of brightnesses. If the input is too dark, the FPA will be unable to properly measure its power spectrum. Conversely, if the input is too bright, it will saturate the FPA. Additionally, the debiased JPS must be rescaled in order to project what would otherwise be negative values on the SLM. While there are various ways to perform this rescaling, it is necessary to ensure that no significant computation time-cost is incurred, and so a fixed rescaling it preferred. The scaling factor is selected by performing a series of correlations with known reference inputs that have typical parameters. The maximum and minimum values for the resulting debiased JPS's are then averaged to give typical upper limits (ULs) and lower limits (LLs). The LL is used as an offset value, such that after rescaling it corresponds to a pixel value of 0, and the UL is used to normalize to a pixel value of 255. This can be summarized as follows:

$$S_{rs} = (S - LL)\frac{255}{UL-LL}$$
$$S_{SLM} = \begin{cases} 0 & S_{rs} \leq 0 \\ 255 & S_{rs} \geq 255 \\ S_{rs} & otherwise \end{cases} \quad (4)$$

where $S_{rs}$ is the rescaled version of $S$, and $S_{SLM}$ is the signal that is sent to the SLM. Because UL and LL are pre-defined values, this computation can be performed on a pixel-by-pixel basis at high-speed using specialized hardware such as a GPU or an FPGA without causing a significant delay in the system. The detailed characteristics of this hardware depend on the precise components being used but can be estimated using the desired performance parameters. We will consider the case of an image with a resolution of 1920 x 1080 pixels on an 8-bit system. Here, there are a total of 2,073,600 pixels, representing the maximum number of discrete rescaling operations for each frame. As mentioned earlier, the frame-speed of the SLMs, $t_{SLM}$, is typically the limiting factor. Thus, a single FPGA dedicated to performing this task would need to process this number of floating-point multiplications faster than $t_{SLM}$. The commercially available Virtex Ultrascape+ FPGA architecture offers 12-bit multiplication units that can operate at up to 872 MHz, each unit utilizing 141 look-up-tables (LUTs) and 189 flip-flops (FFs) [21]. Note that the multiplication units require higher bit-depth than the SLM in order to implement floating-point precision. This same family of FPGAs can have up to 4,086 and 8,172 dedicated LUTs and FFs, respectively. Assuming a maximum of 80% resource utilization in order to leave 20% available for communication and memory requirements, this allows for 23 parallel multiplication units, taking $\frac{1}{872Mhz} = 1.15\ ns$ to complete an operation. In order to rescale an entire frame, this would take $1920 \cdot 1080 \cdot \frac{1}{23 \cdot 872 \cdot 10^6} = 103.4\ \mu s$. Current OJTCs have been demonstrated operating at up to 720 fps using high-speed SLMs [18], yielding a value of $t_{SLM} \approx 1.4\ ms$. Thus, such an FPGA implementation would be significantly faster than required, permitting the use of lower-end FPGAs or alternative devices.

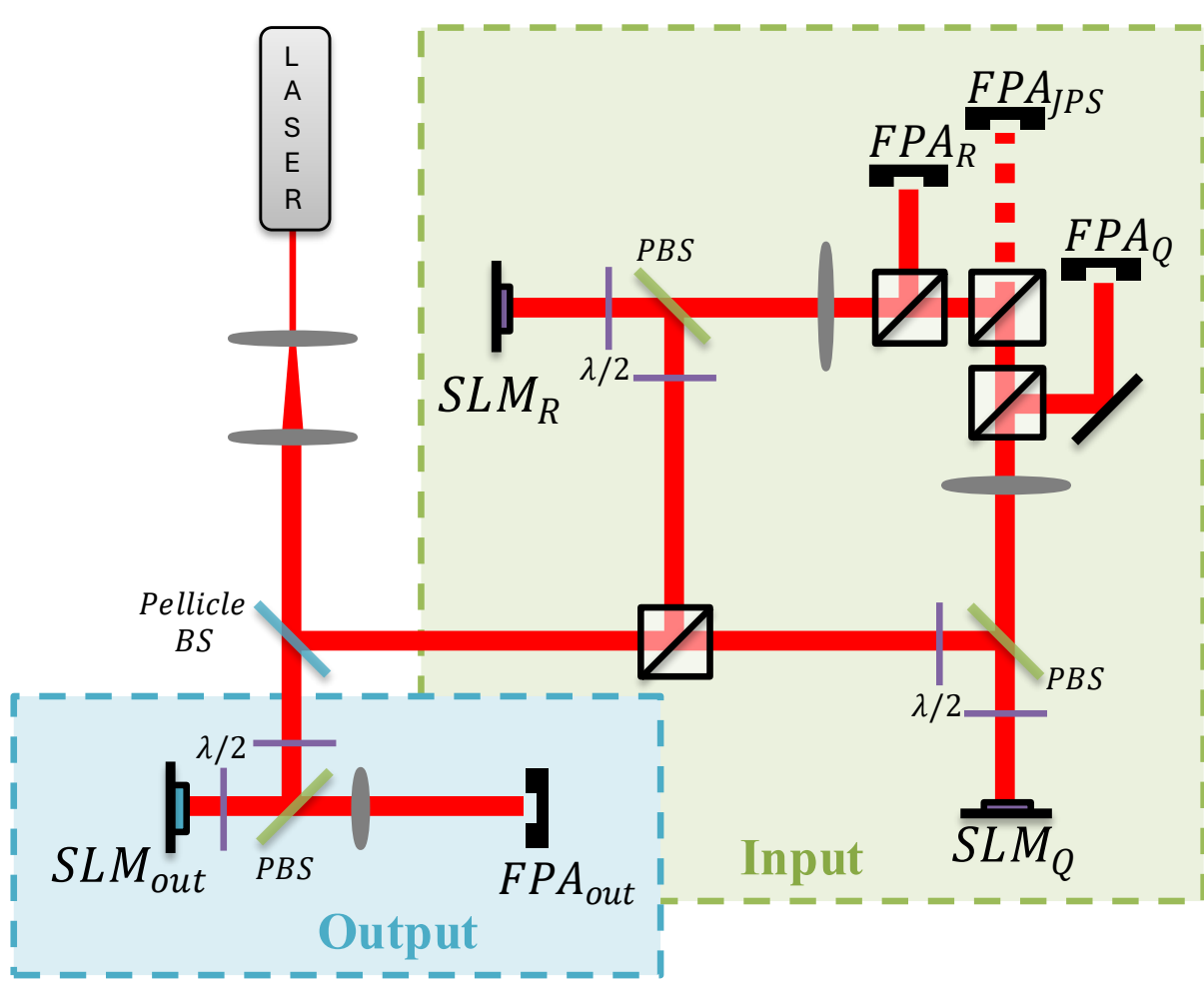

Fig. 3 DOJTC architecture. The reference and query images are projected on $SLM_R$ and $SLM_Q$, respectively. They are separately FT'd using identical lenses, and detected on a combination of FPAs. $FPA_R$, $FPA_Q$, and $FPA_{JPS}$ detect $|\tilde{R}|^2$, $|\tilde{Q}|^2$, and $|\tilde{R}+\tilde{Q}|^2$, respectively. The debiased and rescaled JPS, $S_{SLM}$, is projected on $SLM_{out}$. $FPA_{out}$ detects the intensity of the FT of this signal, which contains the cross-correlation. BS: Beam splitter. PBS: Polarizing BS. λ/2: half-wave plate. A computer (not shown) connects and synchronizes the electronic components.

A diagram of the optical portion of the DOJTC is shown in Fig. 3. Here, the reference and query images are projected on $SLM_R$ and $SLM_Q$, respectively. Each image has an independent path with its own FT lens. A beam splitter (BS) is used to combine and interfere the FTs, allowing $FPA_{JPS}$ to detect their JPS. Prior to reaching this combining BS, each path has a separating BS that functions as a tap to re-direct the FT-beam towards another FPA. Thus, $FPA_R$ and $FPA_Q$ detect $|\tilde{R}|^2$ and $|\tilde{Q}|^2$, respectively. In this particular design, a mirror is placed between $FPA_Q$ and its tap BS in order to ensure reflection parity with $\tilde{Q}$ at the $FPA_{JPS}$ plane. It is important to ensure that all SLM-FPA paths are as equal in optical path length as experimentally feasible. Additionally, alignment is critical in ensuring proper operation of this device, as will be discussed in section 3. It should also be noted that the SLMs must be positioned such that the JPS forms an off-axis hologram, as this offsets the position of the output cross-correlation and pushes it outside of the DC region [14]. This is not necessary for an OJTC as both images are typically projected on the same SLM and as such do not overlap on the input plane, naturally meeting the off-axis condition.

The device shown in Fig. 3 was used for the experiments reported here. However, a simpler version can be constructed such that the input stage uses only one FPA, one lens, and one SLM, requiring an additional shutter or chopper wheel in front of the SLM. In this alternative design, both of the input images are simultaneously projected on the SLM, as in the OJTC. A high-speed shutter or chopper wheel then alternates between three states: 1) fully open, 2) open for the reference image but closed for the query image, and 3) closed for the reference image but open for the query image. The FPA captures the corresponding intensity of the FT for each of the three states, namely 1) $|\tilde{R}+\tilde{Q}|^2$, 2) $|\tilde{R}|^2$, and 3) $|\tilde{Q}|^2$. For maximum speed, the shutter and FPA must operate at 3x the framerate of the SLM. This alternative design avoids the complexity of aligning multiple Fourier beams and FPAs at the cost of limiting the input image size to half of the SLM's resolution while also requiring a well aligned high-speed shutter. Additionally, if high-speed is not required, then the shutter is not necessary. Instead, the system may be constructed exactly like the OJTC shown in Fig. 1, but operated such that the SLM sequentially projects only the required information for each of the three states. This would limit the operational speed to one third of the SLM's framerate, but can be used with OJTC architectures.

### *2.2 Simulated Results*

A series of simulations were performed for a DOJTC and an OJTC using SLMs with a bit-depth of 8-bits and a resolution of 1920 x 1080. For the DOJTC, $S_{SLM}$ was calculated as described in eqn. (4). For the OJTC, the SLM signals were rescaled such that a reference autocorrelation used the entirety of the dynamic range. Three representative results are shown in Fig. 4, where the peaks have been normalized with respect to the result of a DOJTC auto-correlation.

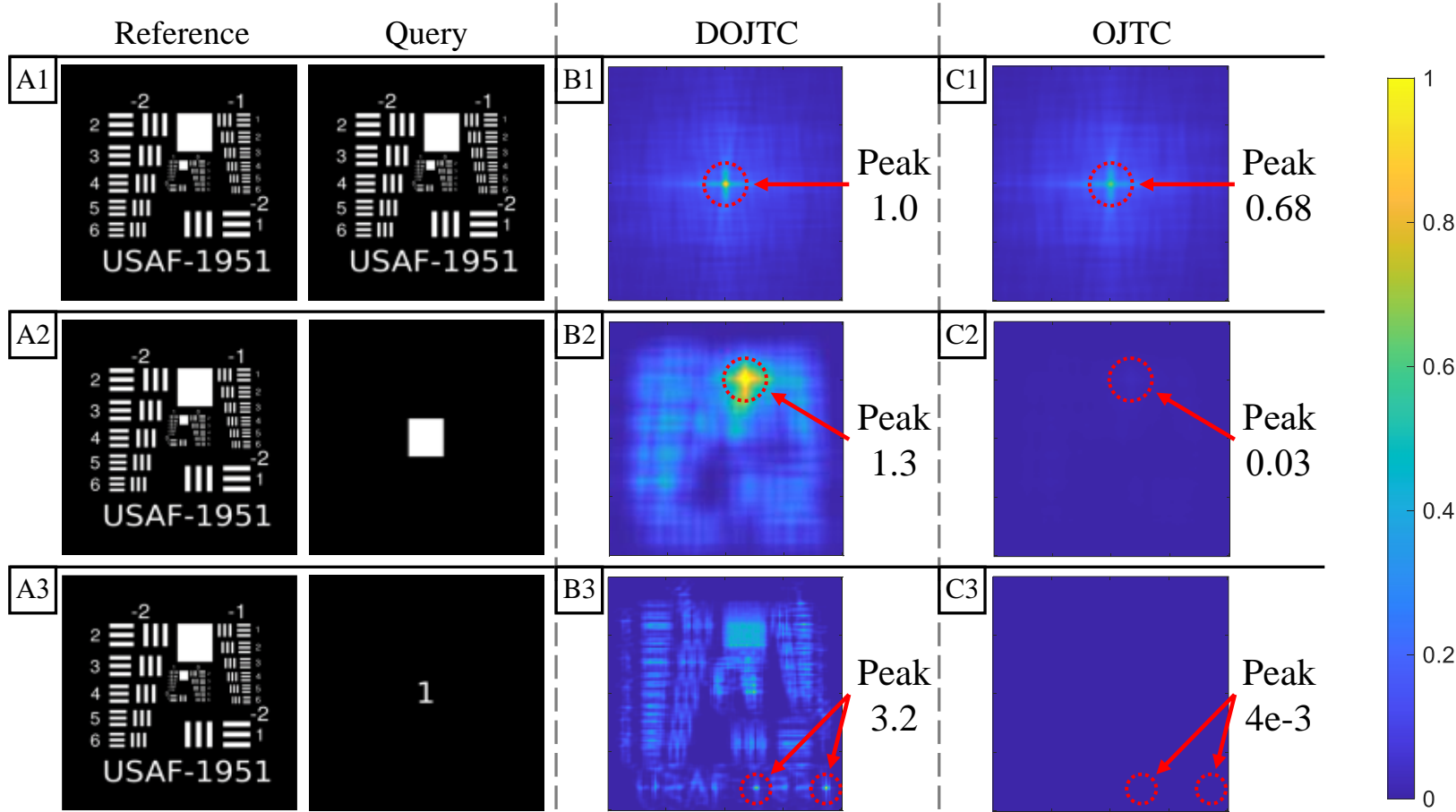

Fig. 4 Simlated results for a seires of correlations. (A): Inputs. (B): Using a DOJTC. (C): Using an OJTC. (Row 1): An autocorrelation of a standard 1951 USAF Resolution Chart (RC). (Row 2): Extracting a square from the RC. (Row 3): Extracting the number '1' from '1951' in the RC. The largest value in each output is highlighted using a red dotted circle and annotated to the right. These values are normalized with respect to the output of (B1).

In the three presented cases, the peaks are always higher in the DOJTC than in the OJTC, showing an improved performance in feature extraction. For the case of an autocorrelation, the OJTC only achieved 68% of the peak value of the DOJTC. In the case of feature extraction, the enhancement was more apparent. When a square is used as the query input, both architectures showed a clear peak, although the DOJTC reached a maximum value of 1.3, whereas the OJTC only achieved 0.03. The drastic decrease in detection power for the OJTC in this case is likely due to the significant difference in total power between the reference and query images. As a result, the $|\tilde{R}|^2$ term dominates in the JPS, resulting in poor utilization of the available resolution by the conjugate product terms. This is further evidenced by the final example in Fig. 4, wherein the query number '1' is to be extracted from the reference '1951.' In this case, the power imbalance results in an even lower peak of $4\cdot10^{-3}$ in the OJTC, whereas the DOJTC achieves a value of 3.2. It should be noted that the DOJTC is able to achieve values greater than 1 thanks to the rescaling, as described in the previous section. In all cases, if the SLMs are assumed to have infinite bit-depths, both the OJTC and DOJTC yield the same results.

## 3. Experimental Results and Discussion

The device shown in Fig. 3 was constructed using ThorLabs Zelux® CS165MU cameras for the FPAs, which have a bit-depth of 10-bits, ensuring that they would not be a limiting factor in these experiments. TI DLP471te modules were used for the SLMs, which have a bit-depth of 8-bits and can be operated at up to 720 fps in monochrome. This SLM has recently been used to construct the fastest OJTC [18].

During operation, the DOJTC measures the JPS of the inputs separately from the independent intensity terms. If the data from each FPA is saved independently, then the device can be operated in one of two modes: either the JPS is directly rescaled and projected on an output SLM without removing the $|\tilde{R}|^2$ and $|\tilde{Q}|^2$ terms, or these terms are subtracted from the JPS prior to rescaling. In the first mode, the device functions exactly as an OJTC, whereas in the second mode it functions as a DOJTC. As such, it is possible to compare the experimental performance of the DOJTC and the OJTC using the same input measurements, as shown in Fig. 5.

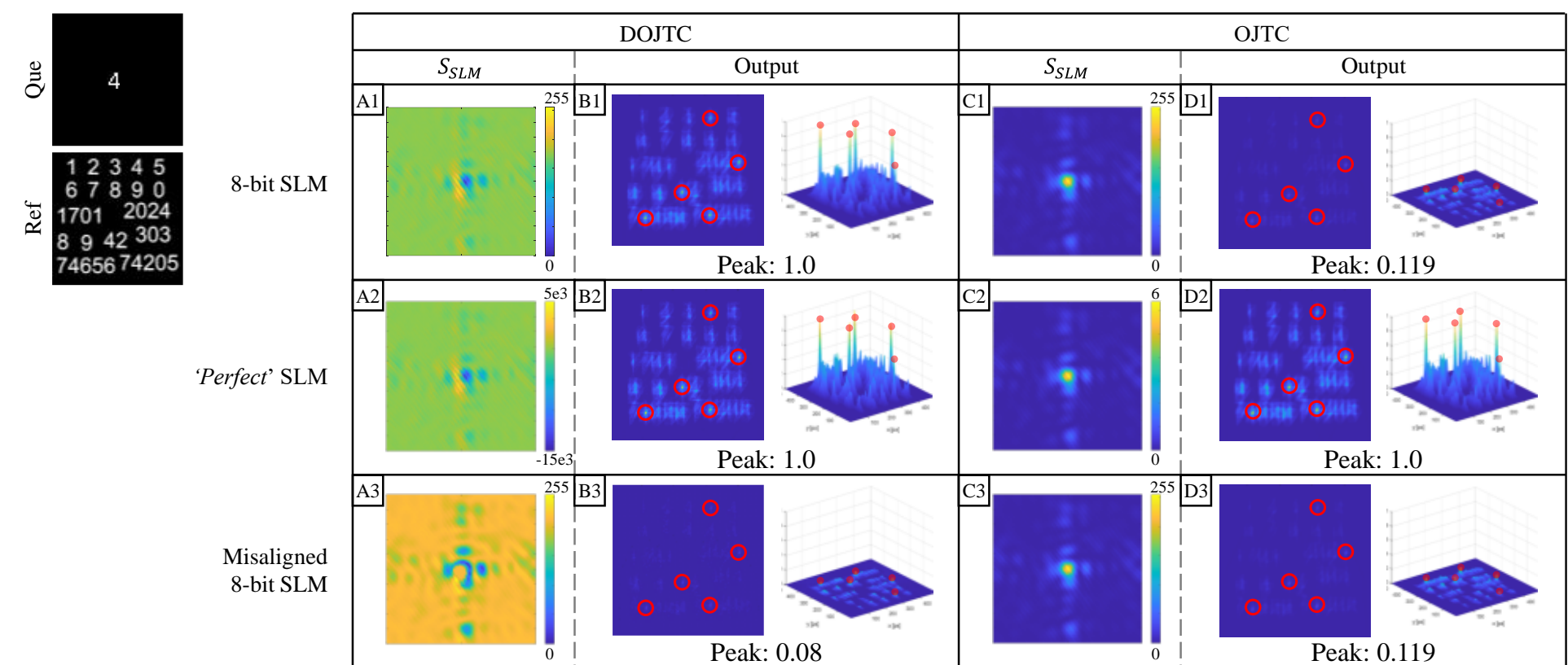

Fig. 5. Experimental results for feature extraction using multiple number strings as the reference input, and the number '4' as the query input. (A): $S_{SLM}$ for a DOJTC obtained from experimental data. (B): Output for a DOJTC. The red circles denote locations of prominent peaks, corresponding to a match. (C): $S_{SLM}$ for an OJTC, corresponding to the experimentally measured JPS from (A). (D): Output for an OJTC. (Row 1): Using an 8-bit SLM to project $S_{SLM}$, and optics to perform the output FT. (Row 2): Using a computer to perform the output FT, corresponding to an SLM with infinite bit-depth. (Row 3): Same as (Row 1), albeit with misaligned inputs.

In the experimental results shown here, the query image was set to be the number '4,' while the reference image was set to be a series of numerical strings which contained five instances of the number '4.' All measurements were normalized with respect to the maximum value of Fig. 5(B1) for simplicity. Fig. 5(A1-D1) shows the case where an 8-bit SLM was used to project $S_{SLM}$ into the optical domain, where it was then FT'd using a lens. Under these conditions, both the DOJTC and the OJTC produced prominent peaks. However, the latter only reached a maximum value of 0.119, yielding a DOJTC enhancement factor of ~8.4. Fig. 5(A2-D2) use the same input data as Fig. 5(A1-D1), but instead perform the output FT on a computer, representing the case where an SLM with infinite bit-depth is used. Here, both the DOJTC and the OJTC reached a peak value of 1. Fig. 5(A3-D3) show the case where an 8-bit SLM was used to project $S_{SLM}$, but there was a misalignment at the input between the JPS and the independent intensity terms measured by either $FPA_R$ or $FPA_Q$. Here, the OJTC produces the same value as in Fig. 5(D1), as the misalignment is irrelevant to its operation. In contrast, the DOJTC's output drops to 0.08, demonstrating how its enhancement is contingent on having appropriate alignment. This could be avoided by using the single-FPA/single-SLM design described at the end of Section 2.1.

The data presented here was obtained at ~30 fps due to the limited framerate of the employed FPAs. However, OJTCs have recently been demonstrated to function at ~720 fps at a resolution of 1920 x 1080 pixels using the same SLMs employed here, albeit with high-speed FPAs [18]. Because the DOJTC's speed is limited in exactly the same way as an OJTC, this same speed can be obtained on the proposed architecture. In contrast, ML based systems are currently limited to ~120 fps at a significantly reduced resolution of ~640x400 [2]. Of course, a direct comparison of speed is insufficient to properly evaluate these two technologies. This is because correlators like the DOJTC are operating over the whole image and producing a mathematical correlation function from which a match can be identified, whereas ML techniques produce an inference through many millions of parameters over billions of small operations. While an in-depth comparison of these two technologies is outside the scope of this paper, we note that recent research has considered the potential of using opto-electronic correlators as large convolution stages for ML algorithms [22].

## 4. Conclusion and Next Steps

Image recognition and convolution are significant parts of our modern world, demanding significant processing throughput to meet the current demand. A wide range of applications could benefit from high-speed correlators. Unfortunately, computers are not optimized for processing 2D data. Image recognition and content matching, for example, currently have to identify markers in order to reduce the amount of data to process. Similarly, convolutional neural networks use a large number of layers, each with small convolutional filters in order to minimize the compute required for a decision. With high-speed correlators, large filters could be processed in parallel, limited only by the resolution and framerate of the employed SLMs and FPAs, which are constantly improving. To this end, opto-electronic correlators can be used as hybrid processing stages, where the JTC is perhaps the most common such architecture. However, the traditional design for an opto-electronic JTC (OJTC) is severely hindered by the self-intensity terms that are unavoidably generated when measuring the JPS of the two input images. For many images, these self-intensities will occupy most of the already limited bit-depth of FPAs and SLMs, thereby compressing the conjugate-product terms to a few digital values and thus reducing the quality of the final result.

Here we have proposed a debiased OJTC (DOJTC) that additionally measures the separate self-intensities of the two images, electronically subtracting them from the JPS to produce a debiased signal that, when projected on an SLM, efficiently occupies the available bit-depth and enhances the result. Simulations show that this can yield an improvement of the normalized peak correlation power of up to two orders of magnitude when compared to an OJTC. Experiments subsequently showed an improvement closer to one order of magnitude when performing feature extraction, demonstrating the real-world benefit of this architecture. As a tradeoff, a strong dependence on optical alignment was observed for the DOJTC, with the result quality dropping drastically with even a minor misalignment. These experiments used a 3-FPA/2-SLM configuration that simultaneously measures all the required signals, but a 1-FPA/1-SLM configuration can be constructed using a chopper wheel or two high-speed shutters. In addition to being simpler, this has the benefit of removing the possibility of misalignment at the cost of decreasing the input SLM resolution by ½ and requiring an FPA that has 3x the framerate of the SLM. As a next step, this simplified version of the DOJTC can be constructed and used for real-time shift, scale, and rotation invariant recognition. Additionally, DMD-based SLMs can increase their framerate by decreasing their bit-depth. Combining this with the proposed DOJTC architecture could enable microsecond-speed opto-electronic correlation with limited impact on the output quality. Further characterization is required in order to find the precise limitations of the architecture using additional performance metrics such as the peak to correlation power [23], signal-to-noise ratio, and energy efficiency.

**Funding.** The work reported here was supported by the Air Force Office of Scientific Research under Grant Agreements No. FA9550-18-01-0359 and FA9550-23-1-0617.

**Disclosures.** The authors declare no conflict of interest.

**Data availability.** Data underlying the results presented in this paper are not publicly available at this time but may be obtained from the authors upon reasonable request.